\documentclass[pra,preprint,aps,showpacs]{revtex4}  

\usepackage{amssymb,amsmath}
\usepackage{graphicx}

\begin{document}

\title{Dynamic electron correlation in interactions of light with matter formulated in $\vec{b}$ space} 

\author{Lev Kaplan}
\email{lkaplan@tulane.edu}
\author{J. H. McGuire}
\email{mcguire@tulane.edu}

\affiliation{Department of Physics and Engineering Physics, Tulane University, New Orleans, LA 70118}

\date{\today}

\begin{abstract}
Scattering of beams of light and matter from multi-electron atomic targets is formulated in the position representation of quantum mechanics.  This yields expressions for the probability amplitude, $a(\vec{b})$, for a wide variety of processes.  Here the spatial parameter $\vec{b}$ is the distance of closest approach of incoming particles traveling on a straight line with the center of the atomic target. The correlated probability amplitude, $a(\vec{b})$, reduces to a relatively simple product of single-electron probability amplitudes in the widely used independent electron approximation limit, where the correlation effects of the Coulomb interactions between the atomic electrons disappear.  As an example in which $a(\vec{b})$ has an explicit dependence on $\vec{b}$, we consider transversely finite vortex beams of twisted photons that lack the translational invariance of infinite plane-wave beams.   
Relatively simple calculations, illustrating the $\vec{b}$-dependence in transition probabilities for photon beams interacting with a two-state degenerate single-electron atomic target, are included.
Further application for many-electron systems is discussed.  Possible practical uses are briefly considered.
\end{abstract}


\maketitle

\section{Introduction}
\label{sec-int}

Physics describes complex objects and processes in terms of simpler ones, successfully to some extent.  There is a wide range of systems composed of light and matter that are well described, both mathematically and conceptually.  However, descriptions of dynamic processes, such as interactions of light with atoms and atoms with atoms, usually depend on understanding the underlying static components.  As a consequence less progress has been made in describing dynamic processes with light and matter that have subsystems of atoms involving more than one active electron, even though most systems we encounter are in this category. In this paper we consider dynamic multi-electronic atomic systems interacting with beams of light and matter.   We begin with the specific example of a vortex beam of twisted photons interacting with a one-electron atom, and extend this to interactions of light and matter with targets that contain more than a single electron.  The variation of twisted vortex beams in the direction transverse to the beam axis leads to an explicit dependence on the translational distance transverse to the beam axis, $\vec{b}$, between the center of the vortex beam and the center of the target. This dependence on $\vec{b}$ is generally absent in descriptions using plane wave photons.

Effects of dynamic electron correlation have been widely observed in interactions of atoms and molecules with both light~\cite{cle1,cle2,cle3,cle4,cle5} and particle~\cite{cme1,cme2,cmet1,cme3} beams.  Of the formulations of the many-body problem available~\cite{cmet1,cme3,mbp,corTh,Lin,Mc}, the one we employ~\cite{Mc} has been used to describe electron correlation dynamics in collisions of multi-electron atoms with charged particle beams, as well as interactions with plane wave photon beams in the few-eV to few-keV regime. 
The widely tested formulation we follow was developed in position space.  Most (but not all) experiments and current applications involving light interacting with atoms~\cite{cle1,cle2} have utilized optical photon beams, such as laser beams, where the wavelength of the photons, $\lambda \sim 5 \times 10^{-7}$~m, is quite large compared to the atomic size, $a_T \sim 5 \times 10^{-11}$~m.  Under these experimental conditions it is somewhat simpler, conceptually and mathematically, to work in a momentum-space representation, describing light as wave-like rather than particle-like, as discussed below.  Nevertheless, in this paper we work in `$\vec{b}$~-space' (position space) rather than `$\vec{q}$~-space'  (momentum space).  In quantum mechanics both representations give the same observable results.
We work in $\vec{b}$~-space here because it follows an available formulation, provides a natural extension of semiclassical methods used in optical texts~\cite{ME}, can be used in interactions involving x-rays, and offers new insight into the nature of quantum dynamics.

\begin{figure}[b]
\includegraphics[width=3in,angle=0]{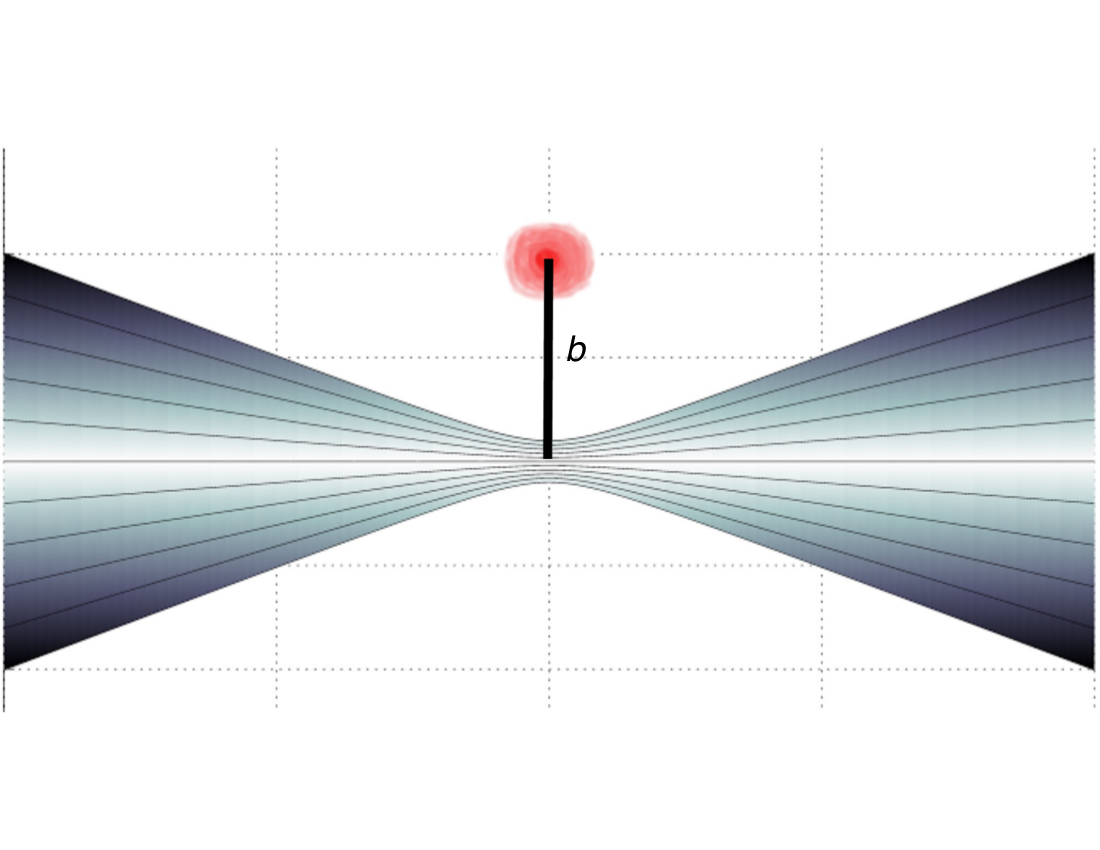}
\caption{
Sketch in the $\hat{b}$-$\hat{z}$ plane of an atom interacting with a Gaussian or Gauss-Laguerre vortex beam~\cite{GKM}.  The maximum of the Gaussian envelope (shown here) of the beam intensity distribution is along the beam axis, and the envelope is cut off when its intensity falls by a factor of $1/e^2 \simeq 0.135$ of its maximum.  This defines the waist size, $w(0)$, of the center of the beam.  The two other independently variable sizes are the mean radius of the atom, $a_T$, and the wavelength of the light, $\lambda$ (not shown here). If the beam is a twisted photon (or electron) beam, it may carry orbital angular momentum, corresponding to a localized photon (or electron) that passes through $\vec{b}$ as it rotates about the $z$-axis. 
The origin of $\vec{b}$ is arbitrary: it may be either at the center of the beam or the center of the atom, for example. 
When the twisted vortex beam carries orbital angular momentum ($\ell \neq 0$), the beam has a more complex geometry~\cite{YaoPadgett}, and has zero intensity along its axis. The atom shown is in its ground state with $\ell = 0$.  In order to exchange angular momentum with a twisted vortex beam with $\ell \neq 0$, the atomic electron must be in an excited state with a non-zero value of $\ell$ that matches that of the twisted photon but has an opposite direction.  In this case the atom has a more complex structure than that shown here and the electronic wavefunction has a node at the center of the atom.
}
\label{Fig1}
\end{figure}

In Sec.~\ref{sec-form}, we formulate electron correlation dynamics in interactions of light and matter with multi-electron atomic systems.  This includes plane-wave beams of light, as well as recently formulated twisted vortex beams~\cite{Allen92,YaoPadgett,Molina-TerrizaTorresTorner}, interacting with single electron atoms~\cite{GKM,DKM}.  The twisted vortex beams are more complex than plane-wave beams; they may carry an orbital angular momentum not present in plane-wave beams.  Moreover the asymptotic angle of the vortex may be adjusted macroscopically and serves as an additional control parameter that affects the interactions of atoms with vortex beams~\cite{DKM,Arot}.  This additional continuously variable parameter is determined by the waist size of the beam vortex, $w(0)$.  Thus we explicitly include three variable-size parameters: the target size, $a_T$, the projectile wavelength, $\lambda$, and $w(0)$.
In Sec.~\ref{calc} we present relatively simple single-electron calculations for photon beams interacting with a two-state degenerate single-electron atomic target.
In Sec.~\ref{sec-disc}, we address some mathematical considerations including the nature of the paraxial approximation~\cite{Allen92} that is often employed for vortex beams.  We also comment on various experimental considerations including the use of our formulation with various targets such as macroscopic gas cells, molecules, and crystals.  Then we address some future applications.  Finally, in Sec.~\ref{sec-sum} we summarize our main results.

\section{Formulation}
\label{sec-form}

In this paper we consider a beam of photons (or electrons) incident on an atomic target, a well-defined initial electronic state $|i\rangle$.  The beam may cause transitions to a particular asymptotic final state $|f\rangle$.  An incoming photon carries momentum $\hbar \vec{k}_i$, while the outgoing photon carries momentum $\hbar \vec{k}_f$.  The momentum transfer is $\hbar \vec{q} \equiv \hbar \vec{k}_f - \hbar \vec{k}_i$.  It is sufficient for our purposes here to consider only elastic scattering where $k_i = k_f$.  This simplifies our notation, and allows us to put aside the effects of a non-zero minimum transfer inessential to this paper, but straightforward to include when needed~\cite{qperp}.  Cross sections and reaction rates for dynamic processes discussed in this paper may generally be described~\cite{cmet1,cme3,mbp,corTh,Lin,Mc,ME} in terms of the scattering amplitude as a function of the momentum transfer, $f(\vec{q})$. 

\subsection{Dual quantum amplitudes in $\vec{b}$~-space and $\vec{q}$~-space}

The equally useful variable, conjugate to $\vec{q}$, is $\vec{b}$.  In this paper we explore uses of the probability amplitude, $a(\vec{b})$, that is conjugate to $f(\vec{q})$.   The physical meaning of $\vec{b}$ itself can depend on the size scale of the projectile compared to that of the target.  In collisions where the beam is diffuse compared to the target, $\vec{b}$ describes the transverse displacement of a point-like atom from the axis of the beam~\cite{GKM}.  When the transverse extent of the beam is small compared to the size of an atom (e.g. in the case of a tightly focused high-energy x-ray beam), $\vec{b}$ describes the transverse displacement of the beam from the center of the atom.  In this paper we generally regard $\vec{b}$ as the transverse displacement between the center of a target and the center of a beam, whose axis is taken as the $z$-axis of the beam-target system.  

The scattering amplitude in $\vec{q}$~-space is related to the probability amplitude in $\vec{b}$~-space by~\cite{MS},   
\begin{eqnarray}
\label{a-f}
  a(\vec{b}) = -\frac{i}{2 \pi  k} \int e^{-i \vec{q} \cdot \vec{b}} f(\vec{q}) \, d\vec{q}   \,.
\end{eqnarray}
In Fourier transforms~\cite{AW1} such as these, if $f(\vec{q})$ is localized in $\vec{q}$, then $a(\vec{b})$ is delocalized and vice versa. 
Both yield the same count rates for physically observable reactions~\cite{AW2}, as illustrated in Eq.~(\ref{cross}) below for the case of total reaction cross sections.  Since $\vec{q}$ is a wave number, Eq.~(\ref{a-f}) may be applied to either classical or quantum wave amplitudes.

Relative size matters.  For optical photon beams interacting with atoms,  $a(\vec{b})$ is generally delocalized compared to the size of a much smaller atom, and $f(\vec{q})$ is localized in $\vec{q}$ space, while for x-rays (or beams of fast electrons or protons), $a(\vec{b})$ may be localized  (i.e., the scattering is approximately particle-like).   In the optical case, $\vec{b}$ describes the location of a well-localized atom within a larger photon beam, whose size is determined by the waist size, $w(0)$, of the beam and the wavelength, $\lambda$, of the photon.  In the case of hard x-ray beams, $\vec{b}$ describes the location of a well-localized middle of the beam trajectory within the atom, whose larger size, $a_T$, is often defined in terms of the Bohr radius, $a_0$.
In any description the physical interpretation of $\vec{b}$ may change as relative size scales change.  In general the size of any system of (possibly overlapping) objects with distinctly different sizes is determined by the size of the largest object.  In scattering of twisted vortex photons with atoms, the distance to which $b$ scales emerges automatically in the scattering amplitude~\cite{GKM} to the larger~\cite{size} of $w(0)$ or $a_T$.

In Sec.~\ref{sec-exp} below, we will address how $a(\vec{b})$ may be used to describe interactions with beams that fall off with distance in the transverse direction from the beam axis, and thus have an explicit dependence on $\vec{b}$.  We also briefly address some aspects of twisted vortex beams of photons and electrons~\cite{GKM}.      
But next we show how $a(\vec{b})$ may be used to to describe interactions of photons with atomic matter in such a way that one may apply previous formulations of electron dynamics to interactions of light with matter.  As a result, cross sections and reaction rates for a larger number of processes may now be calculated, including processes that involve the transition of more than one electron, as well as processes that exhibit the effects of twist in Gauss-Laguerre vortex beams.   

\subsection{Basic formulation of interacting systems in $\vec{b}$~-space}

Since properties of most materials are usually determined by the state of the composite electrons, we seek the dynamic electronic wavefunction, $\psi_{el}$, which may be found by solving the time-dependent Schr\"{o}dinger equation~\cite{Hel},
\begin{eqnarray}
\label{psi}
  H_{el} \psi_{el} = (H_T + H_{int}) \psi_{el} = i \hbar \frac{\partial}{\partial t} \psi_{el} (t) \,.
\end{eqnarray}  

Before the interaction occurs, we assume the electronic state is a known eigenstate $|i\rangle$ of $H_T$, $\psi_{el}(t) =e^{-iE_it/\hbar}|i\rangle$ for $t \to -\infty$.   The interaction, $H_{int}$, changes this state into a superposition of states.  When the interaction has died away, the wavefunction is asymptotically in a superposition of the complete set of eigenstates, namely,
\begin{eqnarray}
\label{psi_f}
\psi_{el}(t \rightarrow +\infty) = \sum_s e^{-iE_s t/\hbar} a_{si}(\vec{b}) |s\rangle \,.
\end{eqnarray}
Using orthonormality of the complete set of basis states, the probability amplitude that the electronic system is in a particular final state, $\langle f|$, is, 

\begin{eqnarray}
\label{a_fi}
  \langle f|\psi_{el}(t \rightarrow +\infty)\rangle = \langle f|\sum_s a_{fi}(\vec{b})|s\rangle = \sum_s a_{si}(\vec{b}) \delta_{fs} = a_{fi}(\vec{b}) \,.
\end{eqnarray}
The observable probability that an electron made a transition from a particular initial state $|i\rangle$, to a possibly different particular final state, $|f\rangle$, is $P = |a_{fi}(\vec{b})|^2$.  
The total cross section for this particular transition is,
\begin{eqnarray}
\label{cross}
  \sigma = \int |a_{fi}(\vec{b})|^2 \, d\vec{b} =\frac{1}{(2 \pi k)^2} \int |f_{fi}(\vec{q})|^2 \, d \vec{q}  \,,
\end{eqnarray}
\noindent
where the last step follows from Parseval's relation for Fourier transforms~\cite{AW2}.
A conventional, straightforward method of evaluating the probability amplitudes is to solve the differential equations~\cite{ME,Hel} arising from Eq.~({\ref{psi}),  
\begin{eqnarray}
\label{a}
  i \hbar \ \dot{a}_{fi}(\vec{b},t) =  \sum_s e^{i E_{fs} t/\hbar} \langle f | H_{int} | s\rangle a_{si}(\vec{b},t) \,,
\end{eqnarray}
\noindent
where $E_{fs} = E_f - E_s$ is the energy difference between the atomic states $|f\rangle$ and $|s\rangle$.
The solutions for the probability amplitudes, $a_{fi}(\vec{b})$,
are often found using the semiclassical approximation for the trajectories of the incoming particles, e.g, $\vec{R}(t) = \vec{b} + \vec{v} t$.  For photons in free space $v = c$.  For a one-electron atom interacting with a photon, $H_T = p^2/2m - Ze^2/|\vec{R} - \vec{r}|$ and $H_{int} = \frac{e^2}{2mc^2} \vec{A} \cdot \vec{A}  - \frac{e}{2mc} ( \vec{p} \cdot \vec{A} + \vec{A} \cdot \vec{p)}$.  Here $\vec{A}$ is the vector potential of the photon field at the location of the atomic electron~\cite{McChap9}, and $\vec{p}$ denotes the momentum operator of the atomic electron.

The formulation above is a standard formulation used for single-electron atomic targets.  This formulation may become useful in some applications where variations in beam flux in directions transverse to the beam axis become significant.  This may include effects where cross sections and reaction rates depend on the distance, $\vec{b}$, of the target from the center of the beam.  

\subsection{Application to multi-electron systems}

Application of this method to interactions of photons with multi-electron targets is straightforward.  
For multi-electron targets, $H_{el}$ becomes~\cite{McChap9},
\begin{eqnarray}
\label{Hel}
H_{el} =  H_T + H_{int} \,,
\end{eqnarray}
  with 
\begin{eqnarray}
\label{HT}
H_T = \sum_{j = 1}^N \left[\frac{ p_j^2}{2m} - \frac{Z_T  \ e^2}{r_j} + e^2 \sum_{k>j} |\vec{r}_k - \vec{r}_j|^{-1} \right] 
\end{eqnarray}
and 
\begin{eqnarray}
\label{Hint}
H_{int} =   \sum_{j = 1}^N \left[ \frac{e^2}{2mc^2} \vec{A}_j \cdot \vec{A}_j  - \frac{e}{2mc} ( \vec{p}_j \cdot \vec{A}_j + \vec{A}_j \cdot \vec{p}_j )\right] \,. 
\end{eqnarray}

\noindent
Here the mass and charge of an electron are denoted by $m$ and $-e$ respectively, $c$ denotes the speed of light, $Z_T e$ is the charge of the target nucleus, $N$ is the number of electrons in the target ($N = Z_T$ for a neutral atom), and $\vec{A}_j$ is the vector potential at the location of the $j^{\rm th}$ electron. 
Thus, the formulation for multi-electron targets~\cite{Mc} is essentially the same as that outlined in Eqs.~(\ref{psi}) -- (\ref{a}) for single-electron systems.  However, detailed calculations become rapidly more difficult as the number of interacting electrons increases~\cite{Kohn}.  
The difficulty, mathematically and conceptually, that arises in solving Eq.~(\ref{psi}) using the multi-electron Hamiltonian of Eq.~(\ref{Hel}) is attributable to the inter-electron interactions, $\frac{e^2}{  | \vec{r}_k - \vec{r}_j |}$, in Eq.~(\ref{HT}).  
In the limit where inter-electron Coulomb interactions may be replaced by a mean field approximation~\cite{meanfield}, i.e.,   $-e^2/|\vec{r}_j - \vec{r}_k| \rightarrow v(r_j)$, the amplitudes $a(\vec{b})$ (as well as the corresponding amplitudes $f(\vec{q})$) for various processes reduce to simple products of single-electron transition amplitudes, and calculations are much easier to deal with.  
Correlation is mathematically characterized by a probability~\cite{cordef,ExInProb} for a process subject to $N \ge 2$ conditions such that $P_{12 \ldots N} \neq P_1 P_2 \ldots P_N$.
That is, only in the widely used uncorrelated independent electron approximation does the probability for an event involving $N$ electrons reduce to a product of single-electron probabilities.

\section{Calculations}
\label{calc}

\subsection{Photons incident on a degenerate two-state atom}

To illustrate our formulation in $\vec{b}$-space, we consider a system consisting of a photon beam interacting with an atom. 
Our photon beam has an electric field given by $\vec{\cal{E}}(x,y,z;t) = \vec{\cal{E}}(\vec{b},z) \cos(2 \pi t/T)$, corresponding to monochromatic light with oscillation period $T$.  The atomic transition involves two states, an initial state $|i\rangle = |1\rangle$ and a final state, $|f\rangle = |2\rangle$.  We focus on events where the state of the atom changes, i.e., $|f\rangle \neq |i\rangle$.  Examples of such dynamic systems include a plane-wave beam, a plane-wave beam with a Gaussian envelope, or a twisted vortex photon incident on an atom, which undergoes a transition involving an exchange of orbital angular momentum with the beam, e.g. a $2s-2p$ atomic transition involving an exchange of angular momentum with the photon.   As needed, one may employ the paraxial approximation so that in the scattering region the light beam is approximately parallel to the beam axis, $\hat{z}$, and the intensity, which may vary with $\vec{b}$, is independent of $z$. 

 Then Eqs.~(\ref{a}) become,
\begin{eqnarray}
\label{atsEq}
  i \hbar \ \dot{a}_{11}(\vec{b},t) &=&  E_1 \;  a_{11}  +   H_{12}(\vec{b}) \cos(2 \pi t/T) \; a_{21}(\vec{b},t)  \nonumber \\
  i \hbar \ \dot{a}_{21}(\vec{b},t) &=&  E_2 \;  a_{21} +   H_{12}(\vec{b}) \cos(2 \pi t/T)  \; a_{11}(\vec{b},t) \,,
\end{eqnarray}
\noindent
where  $   \langle 2 | H_{int} | 1 \rangle  =  \langle 1 | H_{int} | 2 \rangle = H_{12}(\vec{b}) \cos(2 \pi t/T)  $.   

The interaction operator, $H_{int}$, may assume various forms.  For photo-annihilation by optical photons, $H_{int} = - \sum_j \frac{e}{2mc} ( \vec{p_j}\cdot\vec{A_j}  + \vec{A_j}\cdot\vec{p_j})$.  
Then $H_{12}(\vec{b}) = - z_{12} \vec{\cal{E}}_0(\vec{b}) \cos(2 \pi t/T)$, where $z_{12}$ is the dipole matrix element of the atomic transition.  For Compton scattering by x-rays, $H_{int} =  \sum_j \frac{e}{2mc} \vec{A_j}\cdot\vec{A_j} $ and the matrix element, $H_{12}(\vec{b})$, includes higher multipole components, and is related to that of scattering by high energy electrons and protons~\cite{dipole}.   The vector potential, $\vec{A}$, is linearly related to the electric field, $\vec{\cal{E}}$, of the photon beam, and the beam intensity, $I(\vec{b})$, is proportional to $|\vec{\cal{E}}|^2$.    

As a specific example, we now consider the degenerate limit in which $E_1 - E_2 \rightarrow 0$.   In this limit Eqs.~(\ref{atsEq}) have algebraic solutions~\cite{SM}, namely,  

\begin{eqnarray}
\label{ats}
 a_{11}(\vec{b},t) &=&  \ \cos[  \sin(2 \pi t/T) H_{12}(\vec{b})  T /h]  \nonumber \\
 a_{21}(\vec{b},t) &=&  i \sin[     \sin(2 \pi t/T) H_{12}(\vec{b})  T/h ] \,. 
\end{eqnarray}
\noindent 
We have chosen the atomic energy $E_1$ as the zero-point energy of the system.   

\begin{figure}[ht]
\includegraphics[width=3in,angle=0]{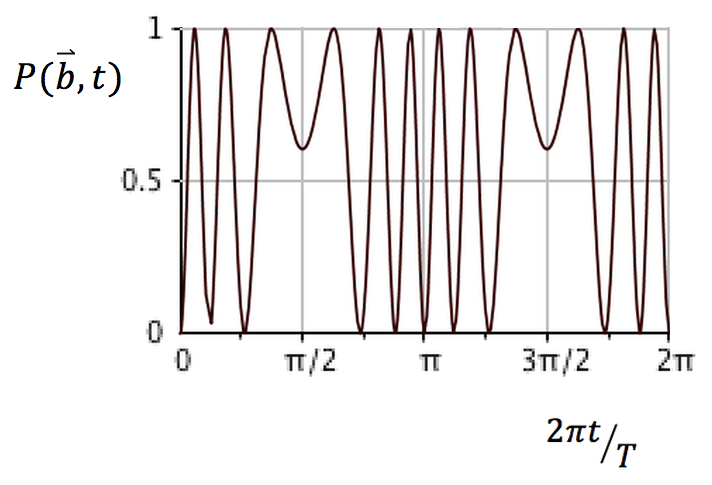} 
\caption{
Probability $P(\vec{b},t)$ of a transition from state $|1\rangle$ to state $|2\rangle$ in a degenerate two-state atom interacting with a plane-wave photon beam is calculated as a function of time, $t$, at a typical value of $\vec{b}$. 
Here $t$ varies from 0 to $T$, where $T$ is the period of oscillation of the photon's electric field; this pattern repeats for longer times. 
In this figure the interaction is non-perturbative: $H_{12}(\vec{b}) T / h = 2.718$. 
}
\label{Fig2}
\end{figure}

The probability for a transition from $|1\rangle$ to $|2\rangle$ is $P(\vec{b},t) = |a_{12}(\vec{b},t)|^2$.   In Figs.~\ref{Fig2}  and \ref{Fig3} we plot this probability at a fixed $\vec{b}$ as a function of time.   The plot shown in Fig.~\ref{Fig2} is a typical result when $H_{12}(\vec{b}) T/h \geq \pi/2$, so that the interaction is non-perturbative and the transition probability may reach unity. In the non-perturbative regime the presence of many, often complex, oscillations is common.  An example where the probability never reaches unity is shown in Fig.~\ref{Fig3} (dashed curve), where $H_{12}(\vec{b}) T / h =1/2$.

\begin{figure}[ht]
\includegraphics[width=3in,angle=0]{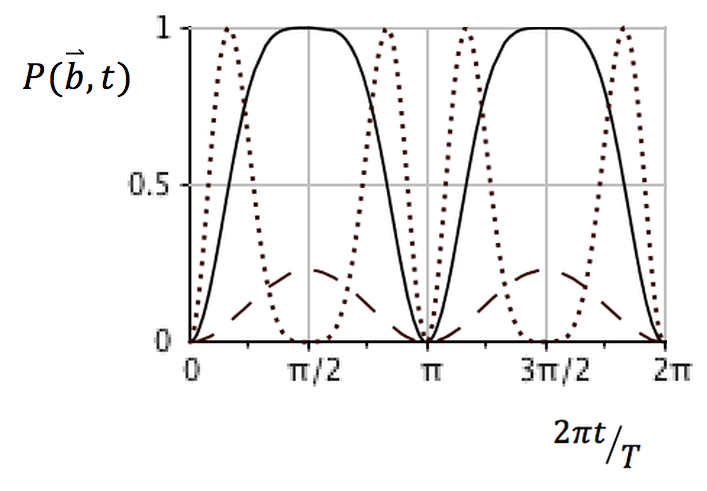}  
\caption{
Probability, $P(\vec{b},t)$, as a function of time, $t$, at special values of $\vec{b}$. 
When $H_{12}(\vec{b}) T / h \geq \pi/2$, the interaction is non-perturbative and the probability can reach unity.
Dashed curve: $\vec{b}$ is chosen such that $H_{12}(\vec{b}) T / h=1/2$. 
Solid curve:  $H_{12}(\vec{b}) T / h =  \pi/2$; here the peak is broad in time~\cite{SM}.
Dotted curve: $H_{12}(\vec{b}) T / h = \pi$; here the probability oscillates regularly, but without the broad maximum in time for the full transfer of the atomic electron population from state $|1\rangle$ to state $|2\rangle$.
}
\label{Fig3}
\end{figure}

Special cases with relatively simple oscillations in time are shown  in Fig.~\ref{Fig3}.   In addition to the perturbative case where $P(\vec{b}, t)$ never reaches unity, two special cases are shown. These special cases occur when $H_{12}(\vec{b})T / h$ is an integer multiple of $\pi/2$.   When the integer multiple is even, maximum population transfer from state $|1\rangle$ to state $|2\rangle$ is relatively short lived.  However, if the integer is odd,  $P(\vec{b}, t)$ has a broad maximum~\cite{SM} around $t = n_{odd} T/4$.  

Next we consider the variation of  $P(\vec{b}, t)$ with the impact parameter, $\vec{b}$.  
As explained above, in general the beam intensity,  $I(\vec{b})$, and thus the strength of the interaction, $H_{int}$, varies with $\vec{b}$.  This produces variations in  $P(\vec{b}, t)$ as $\vec{b}$ varies.   In Fig.~\ref{Fig4} we have plotted $P(\vec{b},t)$ versus 
$R(\vec{b})=  H_{12}(\vec{b}) T/h$ 
at $t = n_{odd} T/4$. This illustrates how the transition probability depends on $\vec{b}$ at times that,
for suitably chosen $\vec{b}$,
correspond to long-lasting complete population transfer.  In this figure we have arbitrarily chosen $R(\vec{b}) = 3 \pi/2$ at $b = 0$, so that complete transfer
is attained there. The value of $R(\vec{b})$ may be controlled by adjusting the overall beam intensity, $I(\vec{b})$. In general Gauss-Laguerre beams are neither isotropic in $\hat b$ nor monotonic in $b=|\vec{b}|$, and neither are the atomic electron densities, at least on some size scales.  In Fig.~\ref{Fig5} we show a plot of the beam intensity ratio, $I(b)$, for Gaussian beams, which are isotropic in $\hat{b}$ and monotonic in $b$.

\begin{figure}[ht]
\includegraphics[width=3in,angle=0]{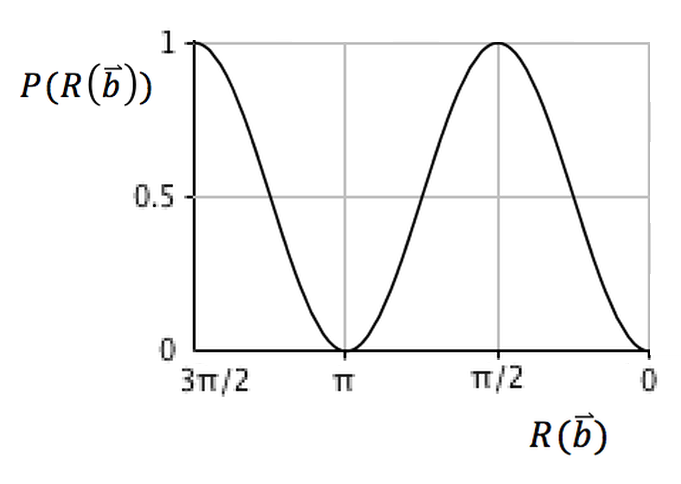}  
\caption{Transfer probability
$P(\vec{b}, t)$ versus $R(\vec{b}) =  H_{12}(\vec{b}) T/h $ occurring at an impact parameter, $\vec{b}$, during an interaction of a Gaussian modified plane-wave photon with a degenerate two-state atom.  The
calculation is performed at $t = n_{odd} T/4$, where $P(\vec{b},t)$ can have a broad maximum in time.  Complete transfer occurs when $R(\vec{b}) = m_{odd} \pi/2$.  
While the horizontal scale in $R(\vec{b})$ decreases from $3\pi/2$ on the left to 0 on the right, 
the impact parameter grows from $b = 0$ to $b = \infty$ since $H_{12}(\vec{b})$ generally falls off with $b$.
}
\label{Fig4}
\end{figure}

\begin{figure}[ht]
\includegraphics[width=3in,angle=0]{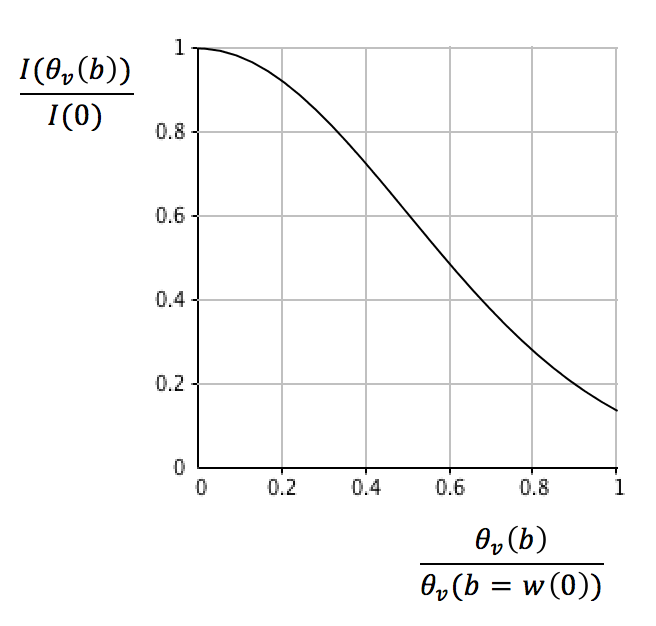}  
\caption{
Intensity distribution of the Gaussian envelope of Fig.~\ref{Fig1} as a function of the angle, $\theta_V(b)$, of the photon trajectory.   $\theta_V(b)$ is the same as the asymptotic cone angle of the beam shown in Fig.~\ref{Fig1}, and is related to the magnitude of the impact parameter, $\vec{b}$, by $\tan\theta_V(b) = b/z_R$, where $z_R$ is the Rayleigh range of the beam.  When $b = w(0)$ the intensity of the beam has dropped by a factor of $1/e^2 \simeq 0.135$ from the maximum intensity at $b = \theta_V(b) = 0$.  Here $w(0)$ is the waist size of the beam, i.e., the beam width at the longitudinal center of the beam, and $w(0) = \sqrt{\lambda z_R /\pi}$ where $\lambda$ is the photon wavelength.  Since $w(0) \ll z_R$ in the paraxial approximation, $\theta_V(b)$ is linearly related to $b$ in the range of interest, and thus the linear horizontal scale in  $\theta_V(b) / \theta_V(b = w(0))$ is equivalent to a linear scale in $b/w(0)$ ranging from 0 to 1.    
}
\label{Fig5}
\end{figure}

In cases when the transition probability is sufficiently small, so that first-order perturbation theory conditions apply,
a perturbative approach using Eqs.~(\ref{ats}) could be used as a basis for 
calculations involving a twisted vortex (Gauss-Laguerre) photon beam interacting with a multi-electron target (including molecules and solids). 
First order perturbation theory is commonly derived~\cite{Hel} using a two-state approximation with the additional condition that the transition probability is always small, i.e., $P(\vec{b},t) = |a_{12}(\vec{b},t)|^2 \ll 1$.
This applies to calculations for both single and double ionization of multi-electron atoms in the high-energy limit under nearly degenerate conditions, where electron correlation effects can dominate multiple-electron transition rates~\cite{McChap9}.

\subsection{Brief overview}

The calculation above for a degenerate two-state atom provides a simple and flexible example illustrating how our formulation works. However, most calculations involving multi-electron effects in atomic scattering are usually much more complex, typically requiring extensive numerical computer codes. Here we give a brief overview of how calculations related to our formulation have been performed.  

We begin with beams of protons, electrons, and ions, which are often particle-like in their collisions with atoms.
Calculations for beams of charged particles interacting with multi-electron targets have followed calculations of various single-electron transitions that can occur in atomic hydrogen including excitation, ionization, and electron transfer (for positively charged particles), in addition to elastic scattering~\cite{Mc}.  For transitions in atomic hydrogen, expressions for $a(\vec{b})$ have been given in sums of closed form in first-order perturbation theory (i.e., first order in $H_{int}$).  These expressions can be used~\cite{Hel} in higher-order perturbation theory, and in coupled-channel calculations based on Eqs.~(\ref{a}).  For a multi-electron atom~\cite{Mc} the static atomic wavefunctions, $|s\rangle$, are generally calculated numerically (e.g. in a Hartree or Hartree-Fock approximation, or as a sum of such terms).  Generally  this improves the accuracy of the calculations of dynamic one-electron transitions for atoms and molecules. To isolate multiple-electron effects, one may calculate cross sections for multiple-electron (often two-electron) transitions.  This often involves use of higher-order perturbation theory.  For high-energy collisions where perturbation theory applies, reliable second-order calculations are now available~\cite{cme3}.  At lower energies, coupled channel calculations are available~\cite{Mc}.  Effects of electrons on partially stripped ionic, or neutral atomic, beams are discussed below.

Fewer calculations of multi-electron dynamics have been done for photon beams than for beams of charged particles.  Calculations with plane-wave photon beams are quite similar to those with charged particles described above, albeit a little less time consuming.  For optical photons these calculations~\cite{McChap9} may be done using the simplifying dipole limit of corresponding calculations for charged particles.  This applies to both single- and multiple-electron transitions. The calculations for multiple-electron transitions are, nonetheless, usually numerical, and time consuming -- especially in the case of detailed calculations for atoms and molecules with many electrons. For x-rays, useful calculations for ratios of double to single ionization have been done by relating charged particle scattering to both photo-annihilation and Compton scattering~\cite{dipole}.  Calculations for beams of twisted vortex photons interacting with atoms and molecules have not yet been done, although an expression for the scattering amplitude, $f(\vec{q})$, has recently been derived for atomic hydrogen~\cite{GKM}.

\section{Discussion}
\label{sec-disc}

\subsection{Mathematical considerations}

In our experience, mathematical expressions for the wave-like scattering amplitude, $f(\vec{q})$, are a little simpler than for the corresponding probability amplitude, $a(\vec{b})$.  On the other hand the probabilities, $|a(\vec{b})|^2$, may be more intuitive to a wider audience, and the unitarity restriction, $|a(\vec{b})|^2 \leq 1$, can be useful in verifying the validity of specific calculations.   To our knowledge there is no formulation of electron correlation dynamics in $\vec{q}$~-space, but we expect it to be straightforward. In the limit of uncorrelated, independent electrons both $f(\vec{q})$ and $a(\vec{b})$ are products of single-electron amplitudes~\cite{prod,cordef,ExInProb}, so the corresponding observables 
are products of one-electron observables. 

We wish to draw attention to the fact that different physical size scales emerge naturally in $f(\vec{q})$ (and consequently in $a(\vec{b})$) when the scales characterizing various parts of the system change~\cite{GKM}.   Thus there is no one scale more fundamental than another in this description.  The parameter, $\vec{b}$, used to locate an object in space, is a chameleon-like mathematical parameter whose physical significance conceptually changes with different relative scales.  

The paraxial approximation~\cite{Allen92} used for twisted vortex photons in our previous paper~\cite{GKM} simplifies the scattering problem by decoupling beam trajectories from the $x$-$y$ plane.
 In this approximation, particle and ray trajectories are approximated as parallel to the axis of the macroscopic beam~\cite{paraxial}, which is taken as the $z$-axis with $\vec{b}$ in the $x$-$y$ plane (as is $\vec{q}$ for forward scattering).  In both the wave and particle limits, the trajectory of a photon may be regarded as a straight line along $z = ct$.  This may also be applied to electron, proton, and some ion beams in the limit that Coulomb scattering of the incident charged projectile with the target can be ignored~\cite{CoulombTraj}.   In the example of transfer of orbital angular momentum between the beam and the target~\cite{GKM}, in this limit the direction of spin of an atom is reversed (like reversing the spin of a boat's propeller) by exchange of the direction of spin with the twisted photon, where the joint photonic-atomic spin axis is the beam axis, which differs in general from the axis of the photon's trajectory~\cite{ell}.  
Mathematical descriptions of twisted vortex beams that avoid the paraxial approximation are available~\cite{MatulaEtal,IvanovSerbo}, but they are more complex both mathematically and conceptually.

\subsection{Experimental considerations}
\label{sec-exp}

Although there presently exist some experimental results on two-electron transitions due to weak interactions of light with few-electron atomic targets, over a range of wavelengths ranging from visible light to x-rays above 10 keV~\cite{cle1,cle2,cle3,cle4,cle5}, many more experiments that detail how multi-electron dynamics works are possible, including experiments using plane waves as well as twisted vortex photons.

As noted at the end of Sec.~\ref{sec-int}, for beams of twisted photons and electrons incident on atomic targets (see Fig.~\ref{Fig1}), there are three size (or distance) scales, $a_T$, $\lambda$, and $w(0)$.  The waist size (minimum beam width) $w(0)$ can be related to another useful parameter by $w(0) = \sqrt{\lambda z_R/ \pi}$, where the Rayleigh range, $z_R$, describes the distance scale on which the vortex beam is approximately parallel to the $z$-axis, i.e., where the paraxial approximation mentioned above is valid. 
The macroscopic beam angle varies with the magnitude of displacement $\vec{b}$, and the Rayleigh range, $z_R$, according to  $\tan\Theta_V(b) = b/z_R$ for Gauss-Laguerre vortex beams~\cite{GKM}.  
Thus, at a fixed value of $z_R$ (and fixed $w(0)$ at a fixed $\lambda$), $\Theta_V(b)$ can be used to macroscopically control cross sections and reaction rates by choosing different $\Theta_V(b)$ within the beam~\cite{2Theta} to vary $b$.  With x-rays this might be used to select specific regions within an atom. In scattering of the beam from the atomic target (see Fig.~\ref{Fig1}), the incoming and outgoing beams, differing by the scattering angle, $\Theta$, share the same impact parameter, $\vec{b}$.  Figure~\ref{Fig1} shows forward scattering at $\Theta = 0$. 

It is possible to do experiments using macroscopic gas cells~\cite{GKM}, so long as the size of the cell along the beam axis, $\Delta z$, is not large compared to the Rayleigh range, $z_R$. That is, the vortex beam need not be focused at the center of the atom so long as the condition required by the paraxial approximation (discussed above) is satisfied~\cite{paraxial}.
Our description generally requires single-collision conditions experimentally, namely that the target be sufficiently diffuse that the effect of scattering from more than a single atom by a single projectile is not significant. 

 Geometric structure factors~\cite{GeoFac} can be used to convert calculations for a single atom to calculations of scattering from targets such as crystals and molecules.   A large number of such structure factors have been calculated in $\vec{q}$~-space and are readily available.  

We point out that a so-called `twist factor' can be used to convert data for beams of plane wave photons to data for twisted vortex photon beams, and could be useful in designing experiments.  This is relatively easy to calculate, although it is presently described in $\vec{q}$~-space~\cite{GKM}.

We note in passing that a virtual impact method has been developed to describe the observed crossover from particle-like to wave-like behavior in collisions of beams of ions carrying electrons scattering from atomic targets~\cite{MMM}.  This involves additional size scales.  The number of such scales grows as the number of electrons on the incoming ion increases.

\subsection{Future applications}

In regard to future applications,
we call attention to the emerging fields of twisted vortex beams~\cite{Molina-TerrizaTorresTorner}, quantum information~\cite{QI}, and quantum control~\cite{Bucks}.  Twisted beams are more complex than plane wave beams, offering new features such as orbital angular momentum and macroscopically adjustable parameters (Rayleigh range~\cite{DKM} and rotational acceleration~\cite{Arot}), which can be used to control transfer of information and reaction rates in interactions of atoms with light and matter. 
Opportunities may also occur in strongly interacting systems, such as beams interacting with atoms and molecules in a regime where $|a(\vec{b})|^2 \simeq 1$, where full control can occur~\cite{Lin,SM,Bucks}.

In this paper we have addressed the description of multi-electron transitions in two-dimensional dual $\vec{b}$ and $\vec{q}$ spaces.  By applying this approach in dual time and energy spaces, it might be possible to interpret recent FAST experiments~\cite{FAST} that probe how quantum processes are both connected and separated, i.e., correlated, in time~\cite{timecor}.

\section{Summary}
\label{sec-sum}

We have mathematically formulated electron correlation dynamics in scattering of light and matter from multi-electron atomic targets by extending an existing formulation for scattering of protons, electrons, ions and plane-wave photons done in a position representation~\cite{Mc} to photon beams that vary (e.g. decrease in intensity) in directions transverse to the beam axis.  The key parameter in this representation is the position, $\vec{b}$, that specifies the minimum distance between the centers of the light beam and of the multi-electron atomic target.  We have presented results of relatively simple calculations that illustrate  $\vec{b}$-dependence in transition probabilities for photon beams interacting with two-state degenerate single-electron atomic targets.  We have more generally discussed interactions of vortex twisted photon beams with multi-electron atomic targets.  Because they are neither monotonic in $b$ nor necessarily isotropic in $\hat{b}$, vortex beams provide a relatively rich dependence on $\vec{b}$ in scattering cross sections and reaction rates in these processes.

\section{Acknowledgments}
 We acknowledge useful discussions with M. Frow,  J. Wolff,  J. Eberly, and Z. Chang.  This work was supported in part by the NSF under Grant PHY-1205788.

{}


\begin{thebibliography}{99}
\bibitem{size}   The wavelength, $\lambda$, is generally smaller than $w(0)$ due to constraints of Heisenberg uncertainty.
\bibitem{cordef}  Effects of particle identity may be incorporated by including symmetrization of wavefunctions for identical particles, such as electrons.  The general definition of correlation we use is described in Ref.~\cite{Mc}, Sec 6.1.
\bibitem{ell}  Orbital angular momentum, $ \vec{\ell} $, could be transferred to atoms by photoabsorption to prepare atomic states well before the exchange of OAM occurs, but it is $\hat{\ell}$, not $| \vec{\ell} |$, that is transferred between the twisted photon and the atom.
\bibitem{2Theta}  We call attention to the difference between $\Theta_V(b)$ and the more widely used `divergence angle of the beam,' $\Theta_{\cal{V}} =\Theta_V(b = w(0))$ that sets the scale for $\Theta_V(b)$, which varies with $b$.  On the axis of the beam photons travel along the axis and there is no beam divergence at b = 0.  As discussed in Ref.~\cite{GKM}, this applies to the incoming or outgoing beam.
\end{thebibliography}
\end{document}